\documentclass[11pt,twoside]{article}
\usepackage{asp2010}
\resetcounters

\begin{document}
\def\gtsima{$\; \buildrel > \over \sim \;$}
\def\ltsima{$\; \buildrel < \over \sim \;$}
\def\gsim{\lower.5ex\hbox{\gtsima}}
\def\lsim{\lower.5ex\hbox{\ltsima}}
\def\Msun{M_{\odot}} 
\def\Zsun{Z_{\odot}}
\def\Zcr{Z_{cr}}

\title{The infant Milky Way} 
\author{Stefania Salvadori$^1$ and Andrea Ferrara$^2$
\affil{$^1$Kapteyn Astronomical Institute, Landleven 12, 9747AD Groningen, NL} 
\affil{$^2$Scuola Normale Superiore, Piazza dei Cavalieri 7, 56126 Pisa, IT}}

\begin{abstract}
We investigate the physical properties of the progenitors of today living Milky 
Way-like galaxies that are visible as Damped Ly$\alpha$ Absorption systems and 
Ly$\alpha$ Emitters at higher redshifts ($z\approx 2.3,5.7$). 
To this aim we use a statistical merger-tree approach that follows the formation 
of the Galaxy and its dwarf satellites in a cosmological context, tracing the 
chemical evolution and stellar population history of the progenitor halos. The 
model accounts for the properties of the most metal-poor stars and local dwarf 
galaxies, providing insights on the early cosmic star-formation. Fruitful links 
between Galactic Archaeology and more distant galaxies are presented.
\end{abstract}

\section{Background}                                        
One of the most popular methods to identify high-redshifts galaxies ($z\approx
2-7$) is by detecting their strong Ly$\alpha$ line. These Lyman Alpha Emitters 
(LAEs) are mainly associated to {\it star-forming galaxies}, and they have been 
extensively used to probe both the ionization state of the Inter Galactic Medium 
and the early galaxy evolution. At lower redshifts, $z < 5$, galaxies with {\it 
high neutral hydrogen column densities}, $N_{HI} > 10^{20.3}$cm$^{-2}$, can be 
identified in the spectra of more distant quasars by means of their strong 
Ly$\alpha$ absorption line. The most metal-poor among these Damped Ly$\alpha$ 
Absorption systems (DLAs), can provide insights on the initial metal-enrichment 
phases of galaxy formation. Recently a DLA with [Fe/H]$\approx -3$ observed at 
$z_{abs}\approx 2.3$, has indeed revealed strong carbon-enhancement and evident 
odd-even effect \citep{cooke2011a}, consistent with the chemical imprint by $Z=0$ 
faint supernovae \citep{kobayashi2011}. Furthermore, all others DLAs with [Fe/H]$ 
< -2$ observed at high-resolution show chemical abundance ratios consistent with 
those of very metal-poor Galactic halo stars \citep{cooke2011b}, thus suggesting 
possible connections between these absorbers and the early building blocks of Milky 
Way (MW) -like galaxies.\\
We determine the physical properties of the progenitors of the MW and its dwarf 
companions by using the merger-tree code GAMETE (GAlaxy MErger Tree \& Evolution), 
which reconstructs the possible star-formation and chemical evolution histories of 
the MW system. The observed Ly$\alpha$ luminosity and Ly$\alpha$ line equivalent 
width are computed using the LAE model by Dayal and collaborators 
\citep{dayal2008,dayal2010} that reproduces a number of 
important observations for high-z LAEs. Adopting the canonical observational criteria 
we identify the progenitors visible as DLAs and LAEs at redshifts respectively equal to 
$z\approx 2.3$ and $z\approx 5.7$. The observable properties of the MW Galaxy and its 
neighboring companions are presented below, from present days back to the time when the 
Universe was only 1~Gyr old.

\section{The Milky Way system at $z = 0$ : Galactic Archaeology}
Very metal-poor stars represent the living fossils of the first stellar generations. 
Their chemical abundance patterns and Metallicity Distribution Functions (MDFs) 
observed in both the stellar halo and in nearby dSph galaxies can provide fundamental 
insights on the properties of the first cosmic sources.
\subsection{First stars and their Cosmic Relics}
Cosmological simulations suggest that first stars formed at $z\approx 15-20$ 
in primordial H$_2$-cooling minihaloes and that were possibly more massive 
than typical stars forming today (Hosokawa et al. 2011). The transition from 
massive to normal stars is expected to be driven by metals {\it and} dust 
cooling, becoming important when the metallicity of the star-forming gas exceeds 
the critical value, $Z_{cr}=10^{-4\pm 1}\Zsun$ \citep{raffa2002}. We use our 
cosmological model to interpret the observed Galactic halo MDF. We find that the 
low-metallicity tail of the MDF strongly depends on the assumed $Z_{cr}$ value 
(Fig.~1). If the observed cut-off \citep{schoerck2009} suggests $Z_{cr}\approx 
10^{-4}\Zsun$, the presence of the four stars at [Fe/H]$<-4.5$ can only be accounted 
if $Z_{cr} < 10^{-5}\Zsun$. In particular, the existence of the most metal-deficient 
star ever, which has {\it total} metallicity $Z\approx 10^{-4.5}\Zsun$ \citep{caffau2011} 
clearly requires $Z_{cr}<10^{-4}\Zsun$. Such a recent discovery definitely proves 
that dust strongly governs the transition from massive to normal stars in the low-Z 
regimes \citep{raffa2002}.
\begin{figure}[h]\hspace{0.5cm} 
  \includegraphics[height=.21\textheight]{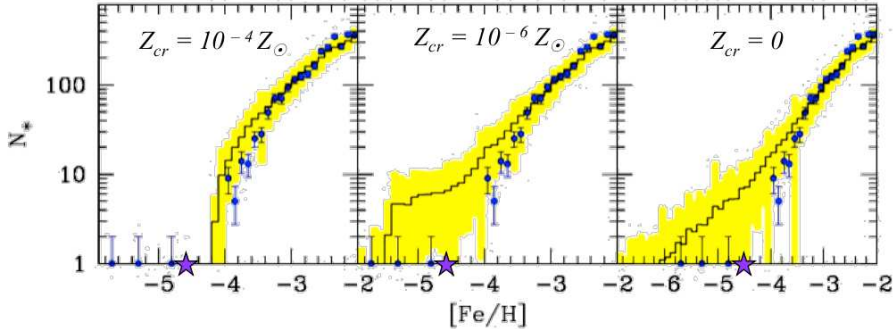}
  \caption{The Galactic halo MDF: observations (points \citep{beers2005}) vs 
simulations averaged over 100 possible MW merger histories (histograms and shaded area). 
The starred symbol indicates the most metal-deficient star.}
  \label{fig:Zcr}
\end{figure}
\\
The chemical abundance patterns of halo stars with $-4.5<$[Fe/H]$<-2.7$
\citep{cayrel2004,caffau2011}, do not 
show any peculiar imprint from very massive primordial stars, and have small chemical 
abundance scatter unlikely resulting from individual supernovae (SN) ejecta. 
According to our cosmological model \citep{salva2007} the number of stars formed 
out of gas polluted {\it only} by $Z < Z_{cr}$ stars is extremely small, and thus 
negligible in current data sample. These "second-generation" stars can either 
have low or high [Fe/H] (Fig.~2) if they form in halos that accreted metal-enriched 
gas from the MW environment or that are self-enriched by the first stars. To have 
the chance to detect the chemical imprint by first stars an higher number of 
[Fe/H]$<-2$ stars is clearly needed. To this aim it is useful to survey the stellar 
halo between $20$~kpc~$\lsim r\lsim 40$~kpc, where the contribution of [Fe/H]$<-2$ 
stars with respect to the overall stellar population is expected to be maximal \citep{salva2010a}. 
\begin{figure}[h]\hspace{0.5cm} 
  \includegraphics[height=.21\textheight]{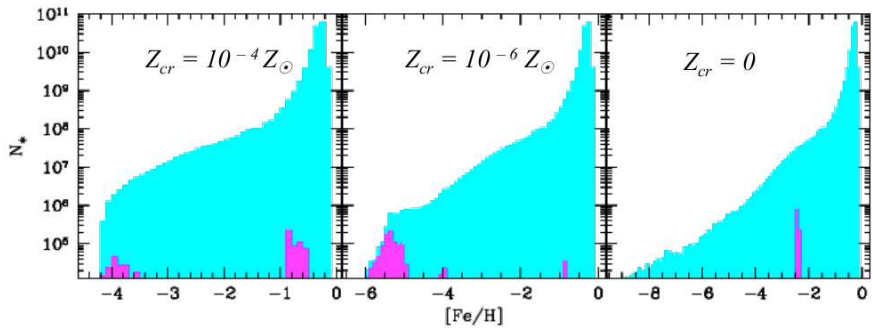}
  \caption{Number of stars predicted to exist at $z=0$ as a function of their 
  [Fe/H] for different $Z_{cr}$ models. The magenta histograms show
  second-generation stars, the cyan histograms the overall stellar populations.}
  \label{fig:2Gs}
\end{figure}
\subsection{First Galaxies and their Cosmic Relics}
An alternative way to find very metal-poor stars is by surveying dSph galaxies
and in particular ultra-faint dSphs ($L < 10^5 L_{\odot}$, Fig.~2), in which 
[Fe/H]$<-3$ stars represent the $25\%$ of the total stellar mass. These faint 
dwarfs are predicted to be among the first star-forming galaxies in the MW 
system, left-overs of H$_2$-cooling minihaloes formed at $z > 8.5$ \citep{salva2009}, 
i.e. before the end of reionization ($z_{rei}=6$). In these galaxies the higher 
fraction of [Fe/H]$<-3$ stars with respect to the more luminous "classical dSphs" 
reflects both the lower star-formation rate, caused by ineffective H$_2$ cooling, and 
the lower metal (pre-)enrichment of the MW-environment at their further formation epoch. 
Indeed classical dSphs are find to finally assemble at $z < 7$ when the pre-enrichment 
of the MW environment was [Fe/H]$\approx -3$. The few stars at [Fe/H]$ < -3$ observed 
in classical dSphs \citep{starkenburg2010} are predicted to form in progenitor 
minihaloes at $z > 8.5$ \citep{salva2009}, some of which might host first stars. 
The unusual composition of two stars at [Fe/H]$\approx -2$ observed in Hercules 
\citep{koch2008} might be the result of self-enrichment by first stars in early 
progenitor halos. 
\begin{figure}[h]\hspace{0.5cm} 
  \includegraphics[height=.20\textheight]{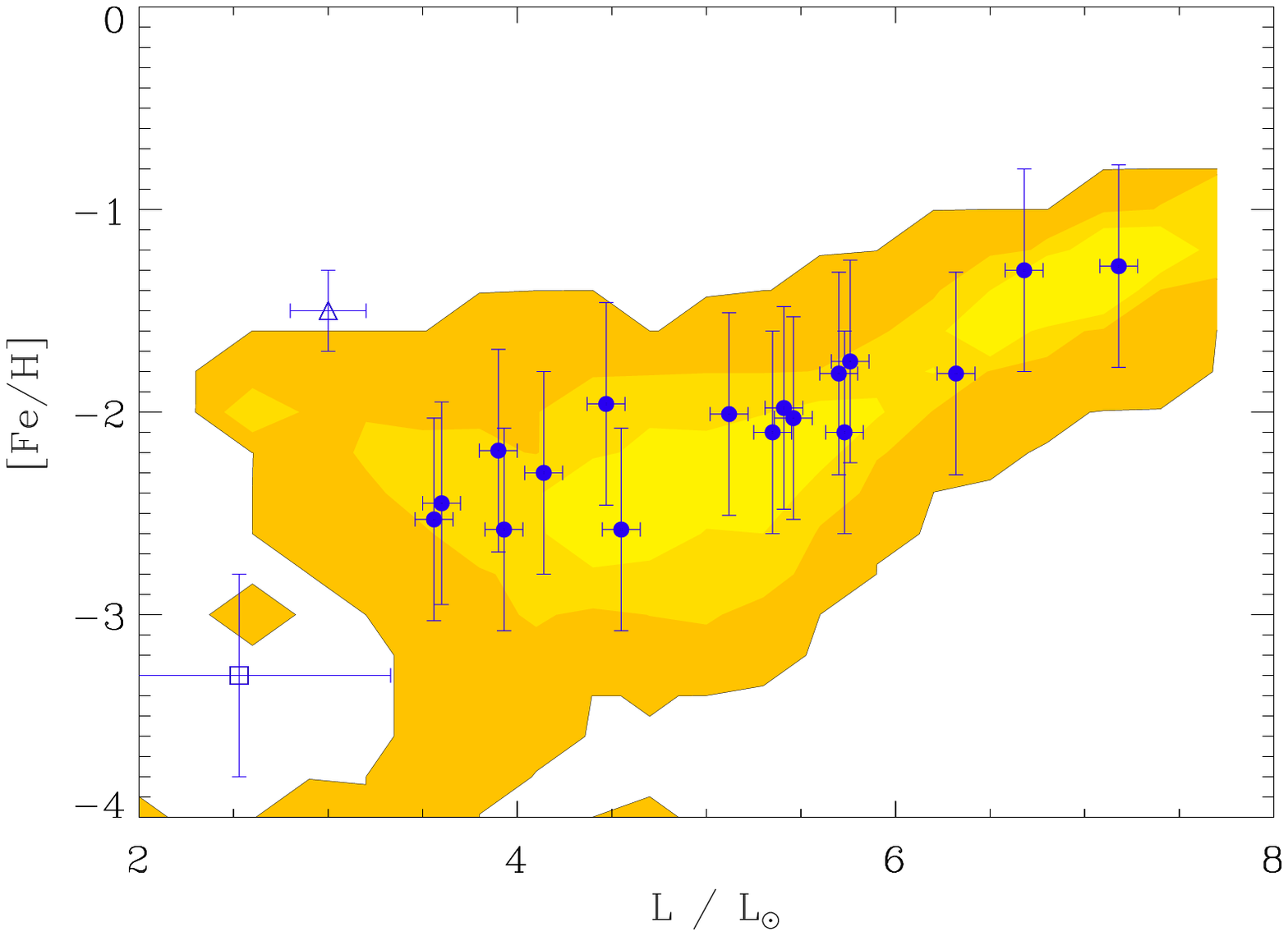}
  \includegraphics[height=.225\textheight]{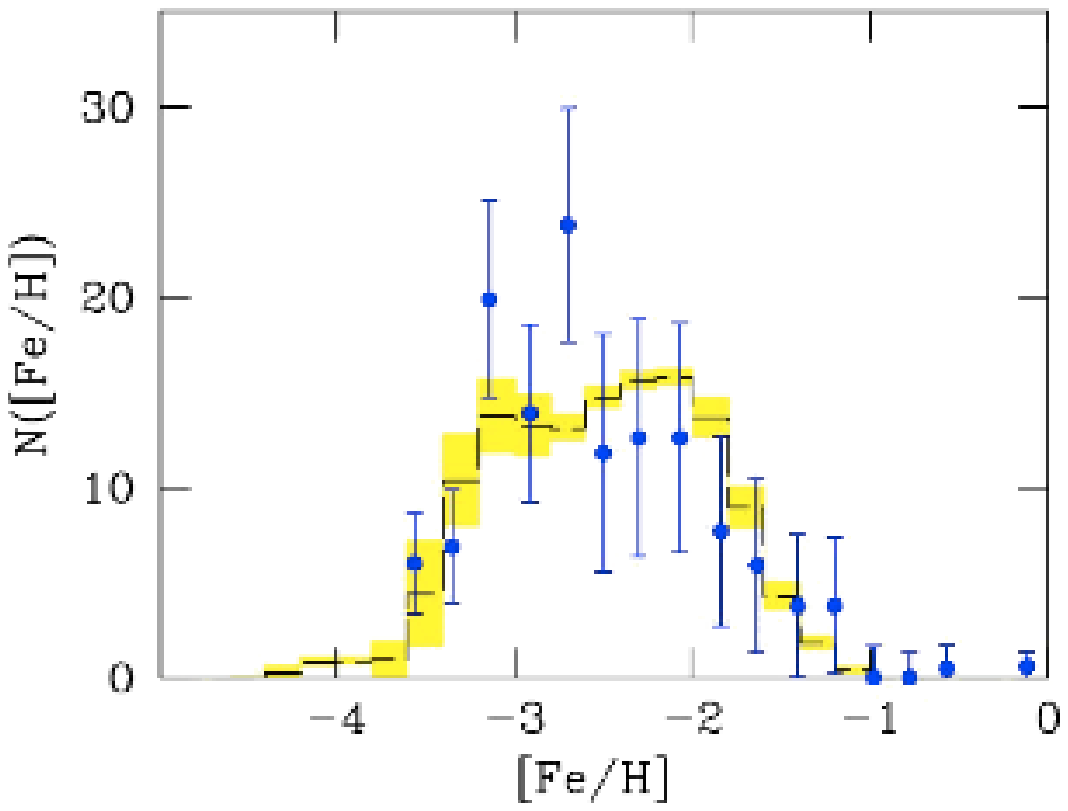}
  \caption{Observed (points with error-bars, \citep{kirby2008}) vs simulated 
(contours/histograms) properties of dSph galaxies at $z=0$. {\it Left:} the 
iron-luminosity relation. The colored shaded areas correspond to regions 
including the $(99,95,68)\%$ of the total number of dSph candidates in 50 
MW merger histories \citep{salva2012}. {\it Right:} MDF of ultra-faint dSph galaxies
\citep{salva2009}.}
  \label{fig:FeL}
\end{figure}
\begin{figure}     
  \includegraphics[height=.21\textheight]{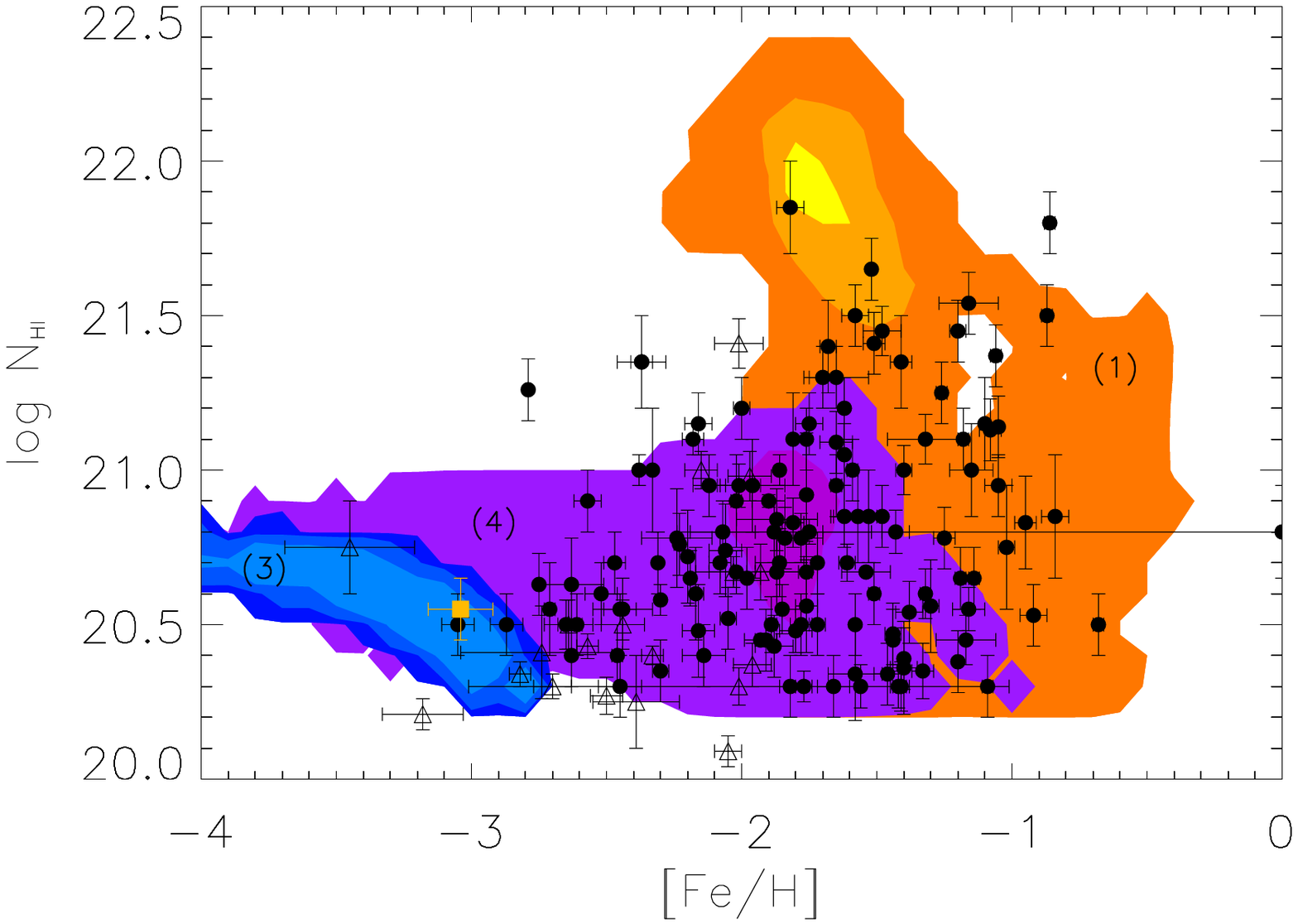}  
  \includegraphics[height=.21\textheight]{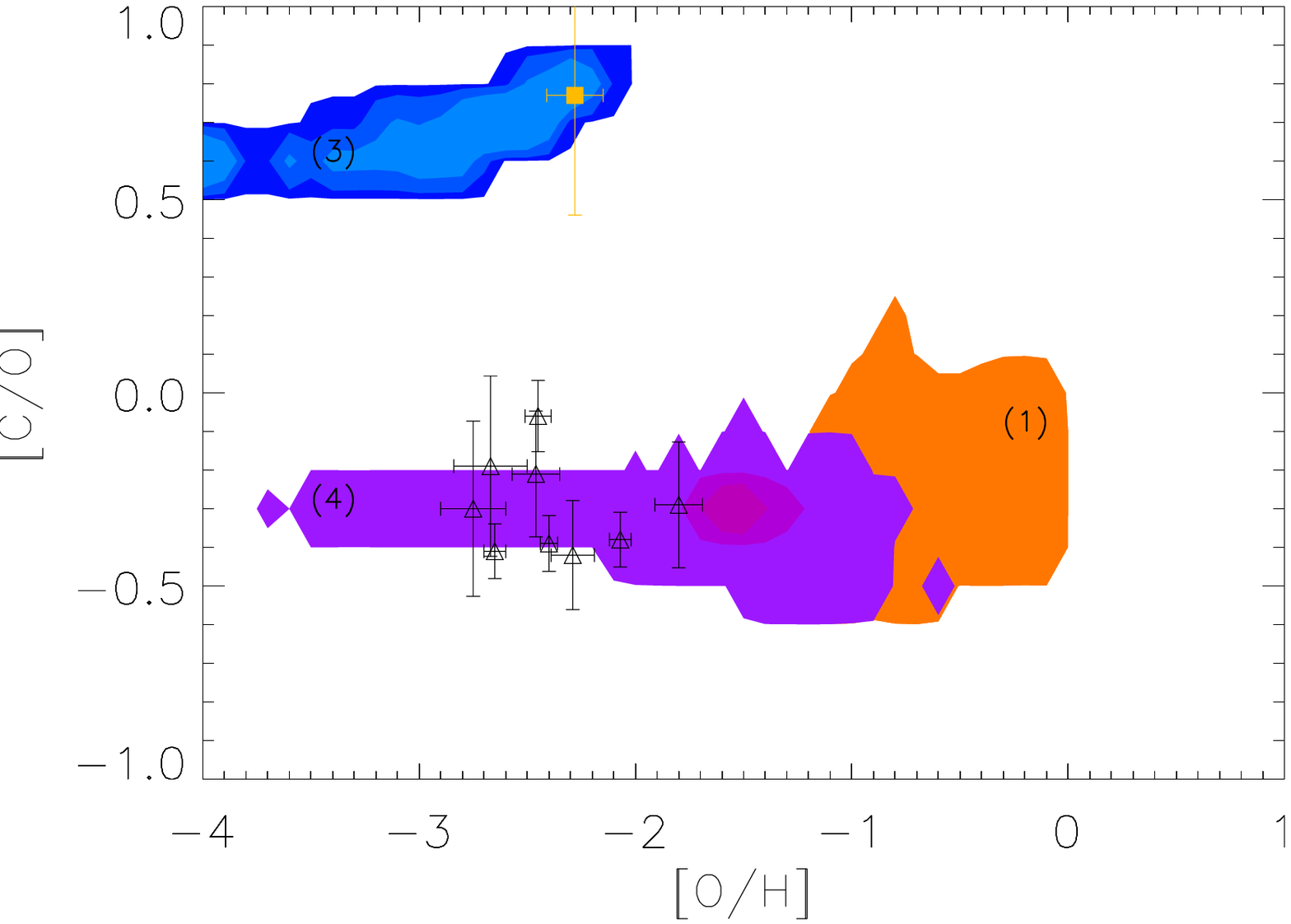}
  \caption{MW progenitors visible as DLAs at $z\approx 2.3$ 
  (contours), with color intensity corresponding to regions containing the 
  ($99,95,68$)\% of DLAs in 50 merger histories \citep{salva2012}. Points with 
  error-bars are observations: circles \citep{prochaska2007}, triangles \citep{cooke2011b},   
  square \citep{cooke2011a}.}
  \label{fig:DLAs}
\end{figure}
\section{The Milky Way system at $z \approx 2.3$: DLAs}
The predicted $N_{HI}$ vs [Fe/H] values of the MW progenitors visible as DLAs 
at $z\approx 2.3$ follows the observed relation (Fig.~4). In our picture very 
metal-poor DLAs, [Fe/H]$<-2$, are associated to starless $M \approx 10^8\Msun$ 
minihaloes that virialize from metal-enriched regions of the MW environment before 
the end of reionization and passively evolve down to $z\approx 2.3$. These sterile 
absorbers retain the chemical imprint of the dominant stellar populations that
pollute the MW environment at their formation epoch: 
low-Z SN type II \citep{salva2007}. This finding agrees with the observational 
results by \cite{becker2012} that show that the gas chemical abundance ratios 
in very metal-poor DLAs/sub-DLAs do not significantly evolve between $2 < z < 5$. 
The recently discovered C-enhanced DLA is instead pertaining to a new class 
of absorbers hosting first stars along with second-generation of long-living 
low-mass stars. These peculiar DLAs are descendants of $M\approx 10^7\Msun$
minihaloes, that virialize at $z > 8$ in neutral primordial regions of the 
MW environment and passively evolve after a short {\it initial period of star 
formation}. These conditions are only satisfied by $\approx 0.01\%$ of the 
total amount of DLAs, making these absorbers extremely rare. The 
peculiar abundance pattern observed in the C-enhanced DLA results from the 
enrichment by low-metallicity SN typeII and AGB stars, which may start to form 
as soon as $Z > Z_{cr}$. While SNII nucleosynthetic products are mostly lost in 
winds, AGB metals are retained in the ISM, causing a dramatic increase of [C/Fe]. 
The amount of N produced by $Z < 5\times 10^{-4}\Zsun$ AGB stars is very limited 
\citep{meynet2002}, resulting in a gas abundance [N/H]$= -3.8\pm 0.9$ (see
Fig.~5). 
The mass of relic stars in C-enhanced DLAs is $M_*\approx 10^{2-4} \Msun$, making
them the gas-rich counterpart of the faintest dwarfs.
\begin{figure}\vspace{0.9cm}\hspace{3.2cm}
  \includegraphics[height=.16\textheight]{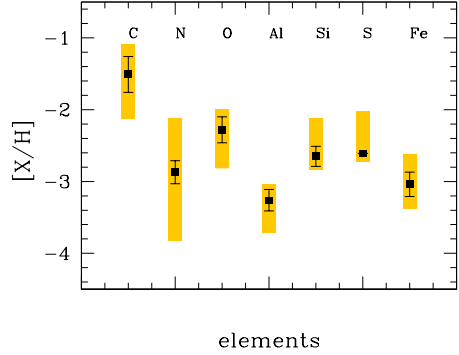}
  \caption{Gas chemical abundances in the C-enhanced DLA: observations (points, 
  \cite{cooke2011b}) vs simulations (average value $\pm 1\sigma$, shaded areas).}
  \label{fig:abundances}
\end{figure}
\section{The Milky Way system at $z \approx 5.7$: LAEs}
At $z\approx 5.7$ the star-forming progenitors of MW-like galaxies cover a 
wide range of observed Ly$\alpha$ luminosity, $L_\alpha = 10^{39-43.25}$~erg
s$^{-1}$. The probability to have at least one progenitor observable as LAE 
($L_\alpha \geq 10^{42}$, $EW \geq 20$ \AA) is therefore very high $P \geq 68\%$. 
Such visible progenitors are mainly associated with the most massive halos of the 
hierarchy, i.e. the major branch, with total mass $M > 10^{9.5} \Msun$. On average 
the identified LAEs have star-formation rates $\dot{M_*}\approx 2.3\Msun/$yr and
$L_\alpha \approx 10^{42.2}$erg/s. They are populated by intermediate age stars, 
$t_*\approx 150-400$~Myr, which have average metallicities $Z\approx (0.3-1)\Zsun$
 (Fig.6). Interestingly these visible MW progenitors provide more than the 
$10\%$ of the very metal-poor stars that are observed today in the Galactic halo. 
Indeed, most of these [Fe/H]$<-2$ stars formed at $z > 6$ in newly 
virializing halos, accreting metal-enriched gas from the MW environment. By $z\approx 
5.7$ many of these premature building blocks have already merged into the major 
branch of the hierarchy, i.e. the visible progenitor \citep{salva2010b}.
\begin{center}
\begin{figure}[!h]\vspace{-0.3cm}\hspace{2.5cm}   
  \includegraphics[height=.35\textheight]{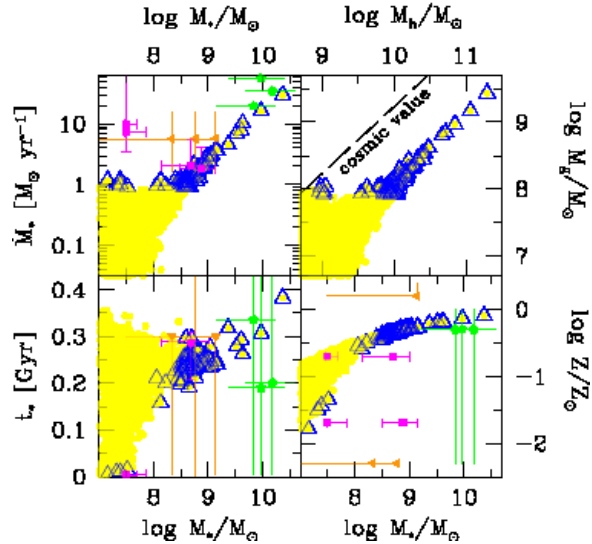}
  \caption{Physical properties of MW progenitors (yellow circles) at $z \approx 5.7$. 
  Blue triangles identify the objects visible as LAEs. Points with error-bars are data 
   by \cite{ono2010} (magenta squares, 4 models for 1 LAE), \cite{pirzkal2007} (blue 
  triangles) and \cite{lai2007} (green circles). As a function of the total stellar mass 
  the panels show: the instantaneous star formation rate (a), the average stellar age (c) 
  and (d) metallicity (d). Panel (b) shows the relation between the halo and the gas mass, 
  with the cosmic value pointed out by the dashed line.} 
  \label{fig:LAEs}
\end{figure}
\end{center}
\section{Conclusions}
The MW system is a powerful laboratory to study early galaxy formation. On the one 
hand the properties of the first cosmic sources can be studied by exploiting the 
observations of today living metal-poor stars and galaxies. From the other hand 
the early evolutionary stages of different MW progenitors can be investigated using 
complementary observations of high-z galaxies. In particular, by identifying the MW 
progenitors among the faintest LAEs observed at $z\approx 5.7$ it will be possible
to observe the MW in its infancy, when it was only 1 Gyr old.
\acknowledgements We thank R. Schneider and P. Dayal for their contribution to the 
works on Galactic Archaeology and MW-LAE connection.

\bibliography{author}
\end{document}